\newcommand{\bld}[1]{\mbox{\boldmath{$#1$}}}

\newcommand{\kalb}{{\textstyle {1\over 2}}}
\newcommand{\beg}{\begin{equation}\label}

\documentstyle[preprint,aps,12pt,floats]{revtex}
\begin{document}
\draft
\tighten

\title{Eight--component two-fermion equations}

\author{Ruth H\"ackl,
Viktor Hund and Hartmut Pilkuhn}

\address{Institut f\"ur Theoretische Teilchenphysik, Universit\"at,
D-76128 Karlsruhe, Germany}

\maketitle

\begin{abstract}{
An eight-component formalism is proposed for the relativistic two-fermion 
problem. In QED, it 
extends the applicability of the Dirac equation with hyperfine interaction to
the positronium case.
The use of exact relativistic 
two-body kinematics entails a CP-invariant spectrum which is symmetric in 
the total cms energy. It 
allows the extension of recent $\alpha^6$ recoil corrections to the positronium
case, and 
implies new recoil corrections to 
the fine and hyperfine structures and to the Bethe logarithm.
\begin{center}{PACS number: 03.65.Pm}\end{center}
}\end{abstract}
\vspace{0.8cm}

\begin{section}{Introduction}\label{ch1}
The relativistic two-body problem for two spin-$1/2$ particles is based on
$16$-component wave functions which transform as the direct product of two
four-component Dirac spinors, $\psi^{(16)} \sim \psi_1  \otimes \psi_2$. For
unequal masses $m_2 > m_1$, the equations are simplified by the elimination of
the small components of particle $2$ and by a subsequent power series expansion
about the non-relativistic limit of this 
particle. One thus obtains an effective
Dirac equation for particle $1$, with a hyperfine interaction that contains the
Pauli matrices $\bld{\sigma}_2$ of particle $2$. Such an equation has $4\times
2 = 8$ components. It is very powerful for hydrogen and muonium
\cite{GY,SY,Petal,Y}.
 More
recently, a non-relativistic quantum-electrodynamics (NRQED) 
has been elaborated which allows one to eliminate the small
components
of both particles, which is particularly useful for the equal-mass case, $m_1 =
m_2$ as in positronium \cite{CL,L}. One thus arrives at an effective Schr\"odinger
equation, in which the Pauli matrices $\bld{\sigma}_1$ and $\bld{\sigma}_2$
produce a four-component spin structure. However, the fact that the power
series expansions in $\alpha$ are rapidly converging, does not prevent
technical difficulties, presently at the order $\alpha^6$ in the binding
energies. These $\alpha^6$-terms have only been calculated by the above
eight-component strategy, and only to first order in $m_1 /m_2$.

In this paper, a different eight-component equation is derived which does not
eliminate small components and dispenses with nonrelativistic expansions. It
exploits the fact that the chirality operator $\gamma_1^5 \gamma_2^5$ commutes
with the matrix $\gamma_1^0 \gamma_2^0$ of the parity transformation; a
corresponding separation of components does not exist for a single Dirac
particle. We begin 
with the rederivation of an eight-component equation for 
two free spin-$1/2$ particles, in which the spin operator of particle 2 is 
removed in the cms ($\bld{p}_1=-\bld{p}_2=\bld{p}$)
 \cite{P92,MP,P95}, by means of a new matrix $c$, given in (\ref{18}) below.
 The spin dependence in the lab system is generated by a
 boost, to be discussed in section 5. The removal matrix 
$c$ also mixes large and small 
components, via a Dirac matrix $\beta=\gamma^0$.
For total cms energy $E$,
the result is an effective single-particle Dirac equation, for a free 
particle of reduced mass $\mu$ and reduced energy $\epsilon$:
\begin{equation}\label{1}\mu=m_1m_2/E,\quad \epsilon=(E^2-m_1^2-m_2^2)/2E,
\quad E=m_1+m_2+E_b.\end{equation}
 In section \ref{3}, the QED Born amplitude for cms scattering will be used 
to derive the interaction for this equation. 
The equation with interaction is for two leptons 
(no anomalous magnetic moments)
\begin{equation}\label{2}
(\epsilon-\mu\beta-V)\psi=\gamma_5(\bld{\sigma}_1-i
\bld{\sigma}^\times V/E)\bld{p}\psi \, , \quad
\bld{\sigma}^\times = \bld{\sigma}_1 \times \bld{\sigma}_2 \, .
\end{equation}
Setting $\gamma_5 \bld{\sigma}_1 = \bld{\alpha} = \gamma^0 \bld{\gamma}$, the
equation has the appearance of the usual Dirac equation with hyperfine
interaction, particularly as $\bld{\alpha} \times \bld{\sigma}_2 /E$ may be
approximated by $\bld{\alpha} \times \bld{\sigma}_2 / (m_1 + m_2 )$ to order
$\alpha^4$. However, the complete operator $ - i \bld{\alpha} \times
\bld{\sigma}_2 V {\bf p} /E$ contains an antihermitian part which ensures exact
relativistic two-body kinematics; for $m_1 = m_2$, it produces the correct spin
structure to order $\alpha^4$ \cite{P95}. A previous derivation from the
$16$-component Dirac-Breit equation produced a different hyperfine operator,
which is equivalent to the new one only near 
threshold,  $-E_b/E\ll 1$. The new hyperfine operator is left 
invariant by the $c$-transformation.
(The previous derivation also had to assume  a point Coulomb potential,
$V=-Z\alpha/r , \, \alpha=e^2, Ze= $ nuclear charge.)

The relativistic on-shell two-body kinematics has long been 
well known. One has 
$\epsilon^2-\mu^2=E_1^2-m_1^2=E_2^2-m_2^2=k^2$ in the cms, where $k^2
=-\kappa^2$ is negative for bound states, and the asymptotic form 
$\rm{e}^{-\kappa r}$ of $\psi$ 
displays $\kappa^{-1}$ as multiple of a Bohr radius. 
The incorporation 
of relativistic two-body kinematics
results 
in a spectrum which expresses $E^2$ 
in 
terms of $m_1^2$ and $m_2^2$. The combination $E\mu=m_1m_2$ is allowed in 
front of a square root (and possibly also with odd powers of $Z\alpha$, 
beginning with $Z^5\alpha^5$). Our main nonperturbative result is
\begin{equation}\label{3}E^2-m_1^2-m_2^2=2m_1m_2(1+Z^2\alpha^2/n^{*2})^{-1/2},
\quad n^\ast = n - \beta_d \, ,
\end {equation}
where $n^{\ast}$ is an effective principal quantum number, and $\beta_d$ is a
 quantum defect. Except for the details of $\beta_d$, (\ref{3}) applies to any
 combinations of spins. It was first derived for two spinless particles 
\cite{Bre}, where also its empirical applicability to parapositronium was 
noticed. It was then extended to the case of one spin-1/2 and one spinless 
particle
\cite{P84}. The angular momentum defect, $\beta_l=-\delta l=-(l'-l)$, has been 
discussed for two  fermions of arbitrary magnetic moments to 
order $Z^4\alpha^4$ \cite{P95}. It will be used in section 8 to derive a 
new ``Barker-Glover'' term
in the fine structure.
In section 4, a rather general formula for $Z^6\alpha^6$ recoil terms will 
be derived which includes several new effects.
In section 6, a vector potential is included in (\ref{2}) and evaluated in 
the dipole approximation. It leads to two additional quantum defects,
\begin{equation}n^*=n+\delta l-\beta_B-\beta'\delta_{l0},\label{3a}\end 
{equation}
where $\beta_B$ is caused by the Bethe logarithm and $\beta'$ is an 
additional quantum defect in s-states. Parts of the Salpeter correction 
\cite{SY}
are not included in (\ref{3a}); their mass dependence is examined in 
section 7. Vacuum polarization and nuclear charge distributions are also 
discussed in that section. For antiprotonic atoms, vacuum polarization must 
be included as a part of $V$ for low-$l$-states. A method is proposed which 
extends the validity of (\ref{3}) to such states. However, also in cases 
where (\ref {2}) must be solved numerically, the $E^2$-dependence remains 
and adds new 
``recoil'' corrections to the binding energies $E_b$ of (\ref{1}).

It may be worth mentioning that the $E^2$-dependence of the spectrum is a very
general consequence of the CPT-theorem. With the separate validity of C,
P, and T in QED, one may also say that the $E^2$-dependence follows from
C-invariance, but in the relativistic case the CP-transformation is slightly
more convenient than C alone. Of course, the states which we calculate are
CP-eigenstates only in the case of positronium. Muonium is transformed into
antimuonium under CP.
\end{section}

\begin{section}{The free eight-component equations}\label{ch2}

Let particles $i (=1,2)$ satisfy the free Dirac equations, in units $\hbar = c
=1$:
\begin{equation}\label{4}
( i \partial_{i}^{0} - \bld{\alpha}_i {\bf p}_i - m_i \beta_i ) \psi_i = 0 \, ,
\quad \bld{\alpha}_i = \gamma_{i}^{5} \bld{\sigma}_i, \quad \beta_{i} =
\gamma_{i}^{0}\, , \quad \partial_{i}^{0} = \partial / \partial t_{i}\, .
\end{equation}
The $\bld{\sigma}_i$ are Pauli matrices, and $\gamma_{i}^{5}\beta_i + \beta_i
\gamma_{i}^{5} = 0$. The direct product $\psi^{(16)}= \psi_1 \otimes \psi_2$
satisfies both equations (\ref{4}) and thus also their sum, in which $i
\partial_{1}^{0}+i\partial_{2}^{0}$ will be replaced by its eigenvalue 
$K^{0}$, which is the total lab energy:
\begin{equation}\label{5}
(K^0 - \gamma_1^5 {\bf p}_1 \bld{\sigma}_1 
- \gamma_2^5 {\bf p}_2  \bld{\sigma}_2
- m_1 \beta_1 - m_2 \beta_2) \psi^{(16)}
= 0 \, .
\end{equation}
$\psi^{(16)}$ is now divided into two octets $\psi_{LP}$ and $\chi_{LP}$, which
have $\gamma_{1}^{5} = \gamma_{2}^{5} \equiv \gamma_{5}$ and $\gamma_{1}^{5} =
- \gamma_{2}^{5} = \gamma_{5}$, respectively:
\begin{eqnarray}\label{6a}
( K^{0} - \gamma_5 {\bf p}_1 \bld{\sigma}_1
- \gamma_5 {\bf p}_2 \bld{\sigma}_2
) \psi_{LP} & = & ( m_1 \beta_1 + m_2
\beta_2 ) \chi_{LP}\, , \\
\label{6b}
(K^{0} - \gamma_5 {\bf p}_1 \bld{\sigma}_1 
+ \gamma_5 {\bf p}_2 \bld{\sigma}_2
) \chi_{LP} & = & (m_1 \beta_1 + m_2
\beta_2 ) \psi_{LP} \, .
\end{eqnarray}
The coupling between $\psi_{LP}$ and $\chi_{LP}$ arises because each $\beta_i$
reverses the eigenvalue of $\gamma_{i}^{5}$. In the chiral basis,
$\gamma_{1}^{5}$ and $\gamma_{2}^{5}$ are diagonal:
\begin{equation}\label{7}
\gamma_{i}^{5} = \left( \begin{array}{cr}
1 & 0 \\
0 & -1 \end{array} \right) , \,\,
\beta_i =
\left( \begin{array}{cc} 
0 & 1 \\
1 & 0 \end{array} \right)  , \,\,
\psi_i = \left( \begin{array}{c} \psi_{ir} \\ \psi_{il}
\end{array} \right) , \,\,
\psi_{LP} = \left( \begin{array}{c}
\psi_{rr} \\
\psi_{ll} \end{array} \right) ,  \,\, 
\chi_{LP} = \left( \begin{array}{c}
\psi_{rl} \\
\psi_{lr} \end{array} \right)  , \,\,
\end{equation}
where the indices $r$ and $l$ (= righthanded, lefthanded) refer to the
eigenvalues $\pm 1$ of $\gamma_{1}^{5}$ and $\gamma_{2}^{5}$. In the parity
basis, the $\beta_i$ are diagonal, with eigenvalues $\pm 1$ for the large and
small components $g$ and $f$, respectively:
\begin{equation}\label{8}
\gamma_i^5 = \left( \begin{array}{cc}
0 & 1 \\
1 & 0 \end{array} \right) \, , \quad 
\beta_i = \left( \begin{array}{cr}
1 & 0 \\
0 & -1 \end{array} \right) \, , \quad
\psi_i = \left( \begin{array}{c}
\psi_{ig} \\
\psi_{if} \end{array} \right) \, ,
\end{equation}
\begin{equation}\label{9}
\psi_{LP} = \left( \begin{array}{c}
\psi_{gg} + \psi_{ff} \\
\psi_{fg} + \psi_{gf} \end{array} \right) \, , \quad 
\chi_{LP} =  \left( \begin{array}{c}
\psi_{gg} -\psi_{ff} \\
\psi_{fg} - \psi_{gf} \end{array} \right) \, .
\end{equation}
In this eight-component space, we also define a new matrix $\beta$:
\begin{eqnarray}
\beta = \beta_{1}\beta_{2} = 
\left( \begin{array}{cr}
1 & 0 \\
0 & -1 
\end{array} \right) \, , \quad 
\beta \gamma_5 + \gamma_5 \beta = 0
\end{eqnarray}
(unit matrices are suppressed). Whereas $\beta_i$ and $\gamma_i^5$ do not
commute, $\beta$ does commute with $\gamma_{1}^{5}\gamma_{2}^{5}$, and both
operators are diagonal in the representation (\ref{9}). As Lorentz
transformations commute with $\gamma_{i}^{5}$ and parity transformations
commute with $\beta_i$, the decomposition of $\psi^{(16)}$ into $ \psi_{LP}$
and $\chi_{LP}$ is invariant under the extended group of Lorentz  ($L$) and
parity ($P$) transformations. 
In the following $\chi_{LP}$ will be eliminated.
 We introduce a compact notation,
\begin{equation}\label{11}
p_{\pm} =  {\bf p}_1 \bld{\sigma}_1 \pm {\bf p}_2
\bld{\sigma}_2 \, , \quad m_{\pm} = m_2 \pm \beta m_1,
\end{equation}
and observe $\beta_2 \psi_{LP} = \chi_{LP} \, ,\quad \beta_2 \chi_{LP} =
\psi_{LP}$ in the basis (\ref{9}), such that we may effectively put $\beta_2 =
1$ and $\beta_1 = \beta$ in (\ref{6a}), (\ref{6b}) in this basis:
\begin{equation}\label{12}
(K^0 - \gamma_5 p_{+} ) \psi_{LP} = m_{+} \chi_{LP} \, , \quad (K^{0} -
\gamma_5 p_{-} ) \chi_{LP} = m_{+} \psi_{LP} \, .
\end{equation}
Using the first equation for the elimination of $\chi_{LP}$, one obtains for 
the second one
\begin{equation}\label{13}
(K^{0} - \gamma_5 p_{-} ) (m_{+} )^{-1} (K^{0} - \gamma_5 p_{+} ) \psi_{LP} =
m_{+} \psi_{LP} \, .
\end{equation}
Multiplying this equation by $m_{+}$ and using 
\begin{equation}\label{14}
m_{+}m_{-} = m_{2}^{2} - m_{1}^{2}\, , \quad m_{+} \gamma_5 = \gamma_5 m_{-},
\end{equation}
one arrives at the following equation
\begin{equation}\label{15}
{\cal K}_0 \psi_{LP} = 0 \, ,\quad 
{\cal K}_0 = 
(K^{0} - \gamma_{5} p_{-} \, m_{-} /m_{+} ) ( K^{0} - \gamma_5 p_{+} ) 
-m_{+}^2 \, .
\end{equation}
Equations (\ref{12}) or (\ref{15}) can be Lorentz transformed to the  cms,
where one has $K^{0} = E ,\,\, p_{-}p_{+}=p_{1}^2 -p_{2}^{2}=0\, .$ The 
constants
of (\ref{15}) are combined into $E^2 -m_1^2 -m_2^2 = 2 E \epsilon$, with
$\epsilon$ defined by (\ref{1}):
\begin{equation}\label{16}
\left\lbrack 2 E \epsilon - E \gamma_5 ({\bf p} \bld{\sigma}
m_{-}/m_{+} + {\bf p} \Delta \bld{\sigma} ) \right\rbrack \psi_{LP} = 2 m_1
m_2 \beta \psi_{LP} , \quad \bld{\sigma} = \bld{\sigma}_1 + \bld{\sigma}_2 \, ,
\quad \Delta \bld{\sigma} = \bld{\sigma}_1 - \bld{\sigma}_2 \, ,
\end{equation}
where $m_1 m_2$ may also be written as $E \mu$ according to (\ref{1}). If ${\bf
p} \bld{\sigma}$ were present without the factor $m_{-} / m_{+}$, one would use
${\bf p} \bld{\sigma} + {\bf p} \Delta \bld{\sigma} = 2 {\bf p}
\bld{\sigma}_1$, and equation (\ref{16}) would be identical with equation
(\ref{2}) for $V=0$, multiplied by $2E$.

It is in fact possible to remove $m_{-} /m_{+}$ from (\ref{16}), by a
transformation
\begin{equation}\label{17}
\psi_{LP} = c \psi \, , \quad c^{-1} \gamma_5 = \gamma_5 c \, , \quad c
\bld{\sigma} c = \bld{\sigma} m_{+} /m_{-}\, , \quad c \Delta \bld{\sigma} c =
\Delta \bld{\sigma} \, .
\end{equation}
Explicit forms of $c$ are
\begin{equation}\label{18}
c = (m_{+} m_{-}) ^{-1/2} \left\lbrack m_2 + \kalb m_1 \beta (1 +
\bld{\sigma}_1 \bld{\sigma}_2 ) \right\rbrack = ( m_{+} m_{-} ) ^{-1/2} ( m_{+}
- 2 m_1 \Lambda_s )\, ,
\end{equation}
where $\Lambda_s = ( 1 - \bld{\sigma}_1 \bld{\sigma}_2 ) /4$ is the projector
on singlet spin states. To verify (\ref{17}), one notices $\Lambda_s
\bld{\sigma} = \bld{\sigma} \Lambda_s =0\, .$ In summary, the $16-$component
equation (\ref{5}) is now transformed into a single free-Dirac equation
$(\epsilon - \mu \beta - \gamma_5 \bld{\sigma}_1 {\bf p} ) \psi_{free} = 0$,
with no trace of the spin operators of particle $2$. In the next section, it 
will be seen that the interaction for this equation contains no mass factors 
at all.

\end{section}

\begin{section}{The one-photon exchange interaction}\label{ch3}

A connection between bound states and perturbative {\sl QED} rests on the
$S$-matrix $S = 1 + iT$ and the Born series for the $T$-matrix, $T = T^{(1)} +
T^{(2)} + \ldots\,\,$. When this series is summed by appropriate differential 
or
integral equations, the bound states appear as poles of $T$. In detail, one
takes plane waves, $\psi_i = u_i e^{i\phi_i}$ with $\phi_i = {\bf k}_i {\bf
r}_i -E_i t$ for the initial states, $\phi '_i = {\bf k}'_i {\bf r}_i - E'_i
t$ for the final states and extracts the resulting energy-momentum conserving
$\delta $-function from the $T$-matrix elements, $S_{if} = i (2 \pi )^4 \delta
(E - E') \delta ({\bf k}_1 +{\bf k}_2 - {\bf k}'_1 - {\bf k}'_2 ) T_{if}$. In
analogy, we put $\psi_{LP} = v e^{i\phi},\,\, \chi_{LP} = w e^{i\phi}$ with 
$\phi =
\phi_1 + \phi_2$. The first Born approximation to $T_{if}^{(1)}$ of the matrix
elements $T_{if}$  is (with $q_1 = -e \, ,\,\, q_2 = Ze \, ,\,\,  e^2 
= \alpha$)
\begin{equation}\label{19}
T_{if}^{(1)} = \frac{4\pi Z \alpha}{t} u'^{\dagger}_1 u'^{\dagger}_2 (1 -
\bld{\alpha}_1 \bld{\alpha}_2 ) u_1 u_2 = \frac{4 \pi Z \alpha}{t} 
\lbrack v'^{\dagger} ( 1 - \bld{\sigma}_1 \bld{\sigma}_2 ) v + w'^{\dagger} ( 1
+ \bld{\sigma}_1 \bld{\sigma}_2 )w \rbrack \, ,
\end{equation}
with $t = {q^0}^2 - {\bf q}^2 \, , \,\, q^0 = K^{0}_1 - {K^0_1}' \, , \,\, {\bf
q}  = {\bf k}_1 - {\bf k}'_1$. Using (\ref{12}) for the elimination of
$v'^{\dagger}$ in one term and of $w$ in the other, $T_{if}^{(1)}$ is expressed
in terms of an $8\times 8$-matrix $M$:
\begin{equation}\label{20}
w = \frac{1}{m_{+}} ( K^{0} - \gamma_5 k_{+} ) v \, , \quad v'^{\dagger} =
w'^{\dagger} ({K^{0}}' - \gamma_5 k'_{-} ) \frac{1}{m_{+}} \, , \quad
T_{if}^{(1)} = w'^{\dagger} M v \, . 
\end{equation}
In the differential equation approach based on (\ref{4}), the potential $V$ is
the Fourier transform of $T^{(1)}$, apart from corrections from the hermitian
part of $T^{(2)}$ \cite{gupta}. Unitarity $S^{\dagger}S =1$ implies
${T^{(1)}}^{\dagger} = T^{(1)}$ and thus $V = V^{\dagger}$, i.e. a hermitian
potential. However, the asymmetric form $w'^{\dagger} M v$ implies a
non-hermitian interaction ${\cal K}_I$ in our differential equation 
$({\cal K}_0 + {\cal K}_I )\psi
= 0$. The simultaneous validity of 
$({\cal K}_0 + {\cal K}_I^{\dagger} ) \chi =0$ guarantees
real eigenvalues, though. The ordinary Dirac equation fails to order 
$Z^4\alpha^4$ in the states with $l = f\; ( l = $ orbital, $f=$ total angular 
momentum), due to hyperfine mixing.

We now restrict ourselves to the cms, ${\bf k}_1 + {\bf k}_2 = {\bf k}'_1
+ {\bf k}'_2 = 0$ and call ${\bf k}_1 = {\bf k}\, , \,\, {\bf p}_1 = {\bf p} =
- i \bld{\nabla}$. The total phase $\phi$ contains ${\bf k} ({\bf r}_1 - {\bf
r}_2 ) = {\bf k} {\bf r}$, such that one has ${\bf p}_2 = - {\bf p}.$
Moreover, $K^0 = {K^{0}}'=E$. To order $Z^4\alpha^4$, one also has $ q^0 = 0 \, , \,\, t = -{\bf q}^2$:
\begin{equation}\label{21}
-{\bf q}^2 M/4 \pi Z \alpha = (E - \gamma_5 k'_{-} ) m_{+}^{-1} ( 1
-\bld{\sigma}_1 \bld{\sigma}_2 ) +m_{+}^{-1} ( 1 + \bld{\sigma}_1
\bld{\sigma}_2 ) (E - \gamma_5 k_{+} ) \, .
\end{equation}
With $k'_{-} = {\bf k}' \bld{\sigma}$, one has $k'_{-} ( 1 - \bld{\sigma}_1
\bld{\sigma}_2 ) = 4 {\bf k}' \bld{\sigma} \Lambda_s = 0$, as $ 1 -
\bld{\sigma}_1 \bld{\sigma}_2$ vanishes for triplet states, while
$\bld{\sigma}$ annihilates the singlet state. Consequently,
\begin{equation}\label{22}
- {\bf q}^2 M/4 \pi Z \alpha = m_{+}^{-1} \lbrack 2 E - \gamma_5 ( 1 +
\bld{\sigma}_1\bld{\sigma}_2 ) {\bf k} \Delta \bld{\sigma} \rbrack = 2
m_{+}^{-1} ( E - i \gamma_5 {\bf k} \bld{\sigma}^{\times} ) \, .
\end{equation}
Combining now $m_{+}^{-1}$ with $w'^{+}$, the Fourier transform of $m_{+} M$
will be called ${\cal K}_I$:
\begin{equation}\label{23}
{\cal K}_I = EV - i \gamma_5 V {\bf p} \bld{\sigma}^{\times} 
\, , \quad V = - Z \alpha /r \,
.
\end{equation}
This is to be 
added to ${\cal K}_0$ (\ref{15}) and
used in (\ref{16}). To arrive at the form (\ref{2}) of the
differential equation, ${\cal K}_I$ must be replaced 
by $c^{-1}{\cal K}_I c$ according to
(\ref{17}). Fortunately, one finds 
\begin{equation}\label{24}
c \bld{\sigma}^{\times} c = \bld{\sigma}^{\times} \, ,\quad c^{-1}
{\cal K}_I c= {\cal K}_I
\, ,
\end{equation}
such that the same operator appears in fact in (\ref{2}).
To order $Z^4\alpha^4$ in $E$, ${\cal K}_I$ 
is equivalent to the operator ${\cal K}_{IB} = EV -
\gamma_5 ( i\bld{\sigma}^{\times} - \bld{\sigma}_2 )\lbrack V , {\bf p}
\rbrack$ of the Dirac-Breit approach \cite{MP,P95}, the transformation being
\begin{equation}\label{24p}
\psi_B =e^{\bld{\sigma}_1 \bld{\sigma}_2 V /2E} \psi \, ,
\end{equation}
after the transformation $c$ (notice $c^{-1} {\cal K}_{IB} \,c \not=
{\cal K}_{IB}$).

For the orders $Z^5\alpha^5$ and $Z^6\alpha^6$, the condition $q^0=0$ should  
be replaced by the current conservation conditions, $q_\mu J_i^\mu=0\;(i=1,2)$.
This produces the spin operator $\bld{\sigma}_1^\perp\bld{\sigma}_2^\perp=
\bld{\sigma}_1\bld{\sigma}_2-(\bld{\sigma}_1\bld{q})(\bld{\sigma}_2\bld{q})
/\bld{q}^2$, which is equivalent to using the Coulomb gauge.
However, we have not yet performed this calculation. 
The new $Z^6 \alpha^6$-terms of (\ref{42a}) below result from the use of
relativistic kinematics in (\ref{2}).
\end{section}

\begin{section}{Solving the new equation nonperturbatively}\label{ch4}

The total angular momentum ${\bf F} = {\bf L} + (\bld{\sigma}_1 +
\bld{\sigma}_2 ) /2 = {\bf J}_1 + \bld{\sigma}_2/2 $ is conserved. All eight
components of $\psi$ have the same eigenvalues $f(f+1) =
F^2$ of ${\bf F}^2$ and $m_f$ of $F_z$; factors $1/r$ and $i/r$ are extracted
from $\psi_g$ and $\psi_f$ as usual:
\begin{equation}\label{25}
\psi_{gk} = \chi_{k}^{fm_f}\frac{1}{r} u_{gk} (r) \, , \quad \psi_{fk} =
\chi_{k}^{fm_f} \frac{i}{r} u_{fk}(r)\, ,
\end{equation}
\begin{equation}\label{26}
(\epsilon - \mu - V) u_g = i {\widetilde \pi} u_f \, ,\quad ( \epsilon + \mu
-V) u_f = - i {\widetilde \pi} u_g \, ,\quad {\widetilde \pi} = (\bld{\sigma}_1
- i\bld{\sigma}^{\times} V /E )r {\bf p} /r \, .
\end{equation}
The index $k$ assumes the four values $k = s = (f,0)$ (singlet), $k = 1  =
(f,1)$ (triplet with orbital 
angular momentum $l = f$) and $k = +\, ,\,\, k = -$
(triplets with $ l\not= f$ and eigenvalues $\mp 1$ of $\sigma_{1r}
\sigma_{2r}=\bld{\sigma_1\hat r\sigma_2\hat r}$) \cite{MP}. We arrange the 
radial wave functions as follows:
\begin{eqnarray}\label{27}
u_g (r) = \left( \begin{array}{c}
u_{g+}\\
u_{g-}\\
u_{g1}\\
u_{gs} \end{array} \right) \, ,\ \quad
u_{f} (r) = \left( \begin{array}{c}
u_{f+}\\
u_{f-}\\
u_{f1}\\
u_{fs} \end{array} \right) \, , \quad
\frac{{\widetilde \pi}}{i} =
\left( \begin{array}{cccc}
0 & 0& -F/r & f_{-} \partial_{-}\\
0 & 0 & \partial_{r} & -f_{-} F/r \\
F/r & \partial_{r} & 0 & 0 \\
f_{+}\partial_{+} & f_{+} F /r & 0 & 0 \end{array}\right) \, ,
\end{eqnarray}
with $\partial_{\pm} = \partial_r \pm 1/r\, , \,\, f_{\pm} = 1 \pm 2V/E$. The
states $k = \pm$ with $ l = f \pm 1$ have $j_1 = j_2 = f \pm 1/2$ to order
$\alpha^2$. An analytic solution of the equations is obtained when the
hyperfine operator is replaced by an equivalent $r^{-2}$-operator. At this
level of precision, the equation becomes equivalent to the one derived from the
Dirac-Breit equation \cite{MP,P95}, where the replacement was achieved by the 
substitution $r=r'-Z\alpha a/2\mu$ (see section 8 for $a$).
\begin{equation}\label{28}
\lbrack \epsilon^2 - \mu^2 - 2 \epsilon V - {\tilde L}^2 / r^2 +
\partial_{r}^{2} \rbrack u = 0.\end{equation}
 ${\tilde L}^2=L^2-Z^2\alpha^2(1+a)$ comprises all $r^{-2}$-operators. Its 
eigenvalues will
be denoted by $l' (l' +1)$, to profit from the analogy with the Schr\"odinger
equation. Defining moreover
\begin{equation}\label{29}
\epsilon^2 - \mu^2 = -\kappa^2 \, ,\quad z = - 2i k r =2 \kappa r
\, ,\quad \alpha_Z=Z\alpha,\quad
\alpha_Z \epsilon /\kappa \equiv n^{\ast}\, ,
\end{equation}
and $ u = e^{-z/2} z^{l'} F(z)$, the equation becomes as usual an equation for
the confluent hypergeometric function $F$,
\begin{equation}
\lbrack l' + 1 - z\partial_{z}^{2} - ( 2 l' + 2 - z)\partial_{z} \rbrack F =
n^{\ast} F \, .
\end{equation}
For unbound states, $\eta = - i n^{\ast}$ is called the Sommerfeld
parameter. For bound states, $ n^{\ast} - l' -1$ must be a non-negative integer
$n_r$. With $l'$ near an integer $l$, one defines
\begin{equation}\label{31}
\delta l = l' -l \, ,\quad n^{\ast} = n_{r} + l' + 1 = n + \delta l \, ,
\end{equation}
where $n = n_r + l+1$ is the principal quantum number and $\delta l$ is always
negative.
All four eigenvalues $\lambda$ of ${\tilde L}^2 - F^2 +\alpha_Z^2$ are given 
by \begin{equation}\label{35}
\lambda^2 \lbrack \lambda^2 - 2\lambda -4 (F^2 -\alpha_Z^2 \epsilon/E ) + 2\alpha_Z^2 \rbrack - 2 \alpha_Z^2 \lambda - \alpha_Z^4 ( 1 - 4
\epsilon/E ) = 0 \, .
\end{equation}
It contains the Dirac eigenvalues with recoil-corrected
hyperfine structure and hyperfine mixing near both static limits \cite{P95}. 
For equal masses, the
equation is reliable only to order $\alpha^2$, but for $l=f$ we nevertheless quote the result for $\delta l$ to order $\alpha^4$. With $1-4\epsilon/E=\alpha^2/4n^2$, one solution of (\ref{35}) is $\lambda=-\alpha^4/8n^2$. It belongs to parapositronium, as we shall see. The value of $l'$ follows from 
$l'+\kalb=\sqrt{F^2+1/4+\lambda-\alpha^2}=\sqrt{(l+\kalb)^2+\lambda-\alpha^2}$,

\begin{equation}\label{36}
-\delta l_{para} = \frac{\alpha^2}{2l +1} + \frac{\alpha^4}{(2 l + 1)^3}+\frac{\alpha^4}{8n^2(2l+1)}.\end{equation}
For orthopositronium with $l = f$, one
factor $\lambda$ is divided off and $\lambda^3$ can be neglected to order
$\alpha^4$: 
\begin{equation}\label{37}
-\delta l_{ortho}(l=f) = \frac{\alpha^2}{2l + 1} \left\lbrack 1-\frac{1}{2F^2}
-\frac{\alpha^2}{2F^6}-\alpha^2 \left(\frac{1-1/2F^2}{2l+1} \right)^2 \right\rbrack-\frac{\alpha^4}{8n^2(2l+1)} .\end{equation}
We now come to our main point, namely the calculation of $E^2$.
From (\ref{29}) and the definition
(\ref{1}) of $\mu$ and $\epsilon$, one finds (\ref{3}). Expansion of the square
root leads to a more practical series, which to order $\alpha_Z^8$ is
\begin{equation}\label{39}
\frac{E^2 -m^2}{m_1 m_2} = - \frac{\alpha_Z^2}{{n^\ast}^2} \left\lbrack 1 -
\frac{3}{4} \frac{\alpha_Z^2}{{n^\ast}^2} + \frac{5}{8} \frac{
\alpha_Z^4}{{n^\ast}^4} \left( 1 - \frac{7}{8} \frac{\alpha_Z^2}{{n^\ast}^2}
\right) \right\rbrack\, ,\quad m = m_1 + m_2 \, .
\end{equation}
Insertion of $n^\ast = n +\delta l$ yields, to order $\alpha_Z^6$,
\begin{equation}\label{40}
\frac{E^2 - m^2}{m_1 m_2} = -\frac{\alpha_Z^2}{n^2}\left(1+\alpha_Z^2
\frac{b}{n}\right) ,\quad b =
-\frac{2\delta l}{\alpha_Z^2}- \frac{3}{4n}+\frac{\alpha_Z^2}{n}
\left(\frac{5}{8n^2}+3\frac{\delta l}{\alpha_Z^2n}+3\frac{\delta l^2}
{\alpha_Z^4} \right). \end{equation}
This expression is still quite compact, in view of the fact that $\delta l$ 
contains both fine and hyperfine interactions. The result for $E$ is, again
to order $\alpha_Z^6$
\begin{eqnarray}\label{41}
E = (m^2 - m_1 m_2\alpha_Z^2 ( 1+\alpha_Z^2b/n) /n^2 )^{1/2}& =&  m - 
\mu_{nr}\alpha_Z^2 (1+\alpha_Z^2b/n)/2 n^2 \nonumber \\ 
&-&\mu_{nr}^2\alpha_Z^4 (1+\alpha_Z^2b/n)^2/ 8 n^4 m - \mu_{nr}^3\alpha_Z^6
/ 16 n^6 m^2,\end{eqnarray}
where $\mu_{nr} = m_1 m_2 /m$ is the non-relativistic reduced mass, and the
expression (\ref{40}) for $b$ remains to be inserted. The third term of
(\ref{41}), with the approximation $b = 0$, is known as the Bechert-Meixner
recoil correction\cite{MP}. To order $\alpha_Z^6$, $b$ may be
approximated by $- 2\delta l/\alpha_Z^2 - 3/4 n$. For comparison with the 
literature \cite{PG,Y}, we split the reduced mass $\mu_{nr}$ in a rather 
unusual way,
\beg{42}\mu_{nr}=\mu_{nr}(1+\mu_{nr}/m)-\mu_{nr}^2/m\approx m_1^2(1-m_1^2/m^2)
-\mu_{nr}^2/m.\end{equation}
In the order $\alpha_Z^6$, we may then combine all contributions from 
(\ref{41}) with that of an additional 
operator $\sim L^2 / r^4$
\cite{PG},
which makes the expression complete for $l>0$:
\beg{42a}\Delta E(\alpha_Z^6)=\frac{\mu_{nr}^2\alpha_Z^6}{2mn^6}
\left(\frac{4n\delta l}{\alpha_Z^2}+1-\frac{\mu_{nr}}{8m}\right)+
\frac{\alpha_Z^2}{2\mu_{nr}^2m}\langle\frac{L^2}{r^4}\rangle.\end{equation}
The last contribution has been symmetrized in the masses, and the 
expectation value $\langle r^{-4}\rangle$ refers to the solution of the 
Schr\"odinger equation with reduced mass. 
This operator has been calculated from two-photon exchange, but the order of
exchange may be gauge-dependent.
Insertion of the Dirac quantum defect
\beg{42b}\delta l_D/\alpha_Z^2=(\gamma-j-\kalb)/\alpha_Z^2\approx -
[1+\alpha_Z^2/(2j+1)^2]/(2j+1)\end{equation} reproduces the known result 
\cite{PG}.
Equation (\ref{42a}) generalizes this result to arbitrary masses and 
hyperfine interactions (section 8).

However, we would like to advocate the direct use
of formula (\ref{40}) for $(E^2 - m^2)/m_1 m_2$, because it is more compact for
calculations and less mass-dependent for measurements.

\end{section}

\begin{section}{boosts and coordinate transformations}\label{ch5}
Lorentz transformations can be constructed directly for $\psi$ 
 and $\psi_{LP}$, but it is more convenient to use the known transformations 
of the single free-particle spinors $\psi_1$ and $\psi_2$. We only need the 
boosts from the  cms to the lab system ($l$), where the system has a 
total four-momentum $K^\mu$:
\begin{equation}\label{42c}\psi_{i,l}=A_i\psi_{i,cm},
\quad A_i=(\gamma+\gamma_i^5
\bld{\widehat K\sigma}_i)^{1/2}=(2\gamma+2)^{-1/2}(\gamma+1+\gamma_i^5
\bld{\widehat K\sigma}_i)\, ,\end{equation}
\begin{equation}\label{43}\bld{\widehat K}=\bld{K}/E,\quad \gamma=K^0/E=(1+
\hat K^2)^{1/2}\, .
\end{equation}
We also take the $z$-axis along $\bld{K},\;\bld{\widehat K\sigma}_i=\hat K
\sigma_{iz}$. Suppressing the index $_{LP}$, the eight-component lab spinor is
\begin{equation}\label{44}\psi_l=A\psi_{cm},\quad A=(\gamma+\gamma_5\hat 
K\sigma_{2z})^{1/2}(\gamma+\gamma_5\hat K\sigma_{1z})^{1/2}=(1+\kalb\hat K^2
\sigma_z^2+\hat K\sigma_z\gamma\gamma_5)^{1/2}\end{equation}
\begin{equation}\label{45}A=\gamma+{\textstyle\frac{1}{2}}\hat K\gamma_5
\sigma_z-\kalb\hat K^2\Delta\sigma_z^2/(2\gamma+2)\end{equation}
(in checking (\ref{45}) by squaring of (\ref{44}), use $\sigma_z\Delta\sigma_z
=0,\;(\Delta\sigma_z)^2=2(1-\sigma_{1z}\sigma_{2z}),\;
(\Delta\sigma_z)^4=4(\Delta\sigma_z)^2$ and $\hat K^4=\hat K^2(\gamma^2-1))$.
The boost $\bar{A}$ for $\chi$ follows from $A$ by replacing $\gamma_5
\bld{\sigma}_2$ by $-\gamma_5\bld{\sigma}_2$:
\begin{equation}\label{46}\chi_l=\bar{A}\chi_{cm},\quad\bar{A}=\gamma+\kalb
\hat K\gamma_5\Delta\sigma_z-\kalb\hat K^2\sigma_z^2/(2\gamma+2).\end{equation}
The inverse boost has $\bld{K}$ replaced by $-\bld{K}$, which is equivalent 
to a sign change of $\gamma_5$:
\begin{equation}\label{47}A\beta=\beta A^{-1},\quad \bar{A}\beta=
\beta\bar{A}^{-1}.\end{equation}
Insertion of (\ref{17}), $\psi_{cm}=c\psi$ gives in the lab system 
$\psi_l=Ac\psi$, and for the $c$-transformed $\psi_l$:
\begin{equation}\label{48}\psi_{lc}=c^{-1}\psi_l=c^{-1}Ac\psi=A_c\psi,\quad
A_c=c^{-1}Ac,\end{equation}
\begin{equation}\label{49}A_c=\gamma+\kalb\gamma_5\hat K\sigma_zm_+/m_--\kalb
\hat K^2\Delta\sigma_z^2/(2\gamma+2).\end{equation}
The corresponding boost for $\chi$, on the other hand, has a factor $m_+$ 
extracted:
\begin{equation}\label{50}\bar{A}_c=m_+^{-1}Am_+=\gamma+\kalb\gamma_5\hat 
K\Delta\sigma_zm_+/m_--\kalb\hat K^2\sigma_z^2/(2\gamma+2).\end{equation}
The desired boosts for $\psi$ and $\chi$ are $A_c$ and $\bar{A}_c$, 
respectively. They are needed for the construction of Dirac-Breit equations 
in the presence of external potentials.

In a covariant treatment, the interaction between two particles at distance $
\bld{r}_l=\bld{r}_1-\bld{r}_2$ depends also on a time difference $x^0=t_1-t_2$,
such that $x^\mu=(x^0,\bld{r}_l)$ is a four-vector. A second independent 
four-vector $X^\mu$ is defined such that $P^\mu=i\partial_X^\mu$ becomes the 
total four-momentum $p_1^\mu+p_2^\mu$, which is conserved:
\begin{equation}\label{51}x^\mu=x_1^\mu-x_2^\mu, \quad X^\mu=\hat E_1x_1^\mu+
\hat E_2x_2^\mu,\quad p_1^\mu=p^\mu+\hat E_1P^\mu,\quad p_2^\mu=-p^\mu
+\hat E_2P^\mu,\end{equation}
\begin{equation}\label{52}\hat E_1+\hat E_2=1,\quad P^\mu=p_1^\mu+p_2^\mu,\quad
P^\mu\psi=K^\mu\psi.\end{equation}
The as yet open value of $\hat E_1-\hat E_2$ is chosen such that $p^0=i\partial
/\partial x^0$ vanishes at asymptotic distances in the cms where particles 1
 and 2 are on their mass shells, $E_1^2-k^2=m_1^2,\;E_2^2-k^2=m_2^2,$ i.e. 
$E_2^2-E_1^2=m_2^2-m_1^2=m_+m_-$. Using in addition $E_1+E_2=E$, one obtains
\begin{equation}\label{53}E_1=(E^2-m_+m_-)/2E,\quad E_2=(E^2+m_+m_-)/2E.
\end{equation}
Extracting now from (\ref{51})\begin{equation}\label{54}p^\mu=\hat 
E_2p_1^\mu-\hat E_1p_2^\mu\end{equation} and inserting the asymptotic values 
$E_i$ of $p^0_i$ in the cms, one finds that $p^0$ vanishes here for 
\begin{equation}\label{55}\hat E_i=E_i/E=\kalb (1\mp\hat m_+\hat m_-),\quad 
\hat m_\pm=m_\pm/E.\end{equation} In the interaction region, $p^0$ does not 
vanish,
but in the context of a single integral equation (Bethe-Salpeter equation) it 
cannot be treated as a dynamical variable. Instead, integrals involving $p^0$ 
are treated as perturbations on $\delta(p^0)$-integrals (section 7).

The space coordinate transformation of (\ref{51}) is
\begin{equation}\label{56}\bld{r}_1=\bld{R}+\hat E_2\bld{r}_l,\quad\bld{r}_2
=\bld{R}-\hat E_1\bld{r}_l,\end{equation} where $z_l$ is Lorentz contracted, 
$z_l=z/\gamma$. The nonrelativistic approximation yields the familiar $\hat 
E_i=m_i/m$.
\end{section}

\begin{section}{Vector potential and Bethe logarithm}\label{ch6}
In the presence of a four-potential $A^\mu,\; p_1^\mu$ and $p_2^\mu $ are 
replaced by
\begin{equation}\label{57}\pi_1^\mu=p_1^\mu+eA^\mu(x_1),\quad\pi_2^\mu 
=p_2^\mu-ZeA^\mu(x_2).\end{equation}
The coefficients $\hat E_1$ and $\hat E_2$ of the transformation (\ref{51}) 
to $\pi^\mu$ and $\Pi^\mu$ remain unchanged in a perturbative treatment of 
$A^\mu$. In the dipole approximation $\bld{A}(x_1)=\bld{A}(x_2)$, (\ref{54})
leads to \begin{equation}\label{58}\bld{\pi}=\hat E_2\bld{\pi}_1-\hat 
E_1\bld{\pi}_2=\bld{p}+re\bld{A},\end{equation}
where $re=r\cdot e$ is the dipole radiation charge \cite{MP},
\begin{equation}\label{59}r=\kalb(Z+1)-\kalb(Z-1)\hat m_+\hat m_-=1
+\kalb(Z-1)(1-\hat m_+\hat m_-).\end{equation}
In the following, we use the Coulomb gauge in the cms and replace $\bld{p}$ by
$\bld{\pi}$ in (\ref{2}). To exhibit its $E^2$-dependence, we also multiply 
the equation by $E$:
\begin{equation}\label{60}(E\epsilon-EV-m_1m_2\beta)\psi=\gamma_5(
\bld{\sigma}_1-i\bld{\sigma}^\times V/E)E\bld{\pi}\psi.\end{equation}
The substitution \begin{equation}\label{61}\bld{r}=E\bld{\rho}, \quad\bld{p}
=\bld{p}_\rho/E,\quad\bld{\pi}=\bld{\pi}_\rho/E\end{equation}
brings (\ref{60}) into a form which contains only even powers of $E$. For the 
moment, we assume $V=-Z\alpha/r,\; EV=-Z\alpha/\rho=V(\rho)$. Deviations will 
be discussed in section 7. For the Lambshift calculation, one neglects the 
hyperfine operator and obtains an explicit eigenvalue equation for $E\epsilon
=(E^2-m_1^2-m_2^2)/2$,
\begin{equation}\label{62}h\psi=E\epsilon\psi,\quad h=m_1m_2\beta +V(\rho)+
\bld{\alpha}\bld{\pi}_\rho.\end{equation}
Its Coulomb Greens function G satisfies the usual equation \cite{SY}, taken 
in the new variables $\bld{\rho,\rho'}$:
\begin{equation}\label{63}[\nabla^2_\rho +(E\epsilon)^2-m_1^2m_2^2+2E\epsilon 
Z\alpha/\rho+(Z\alpha+i\bld{\alpha\hat\rho})Z^2\alpha^2/\rho^2]G(\bld{\rho,
\rho'},E\epsilon)=\delta(\bld{\rho}-\bld{\rho'}).\end{equation}
Evidently, it is also independent of the signs of $m_1$ and $m_2$. The same 
remark applies to $r$ (\ref{59}), but that expression uses the low-energy 
dipole approximation. Apart from that,
 the Bethe logarithm can only depend on $E^2,\;m_1^2$ and $m_2^2$. Its 
proportionality to $r^2$ has been noted previously 
\cite{E,K}. Inspection of the formulas collected in \cite{SY} reveals another 
small s-state correction, which is also proportional to $r^2$. Moreover, both 
corrections are proportional to $m_1m_2$ and therefore pushed under the 
square root in (\ref{3}), where they appear as quantum defects:
\begin{equation}\label{64}\beta_B=\frac{4\alpha^3}{3\pi}r^2\ln\, k_0(n,l),
\quad
\beta'= -\frac{4\alpha^3}{3\pi}r^2\left(\frac{5}{6}-\ln\,\alpha^2\right).  
\end{equation}
After the extraction of $\beta_B $ and $\beta'$, there remains a somewhat 
reduced Salpeter shift, \begin{equation}\label{65}
\Delta E_{Sal}'=-\frac{\mu_{nr}^2Z^5\alpha^5}{m\pi 
n^3}\left[\frac{7}{3}a_n'+\frac
{1}{m_+m_-}\delta_{l0}\left(m_2^2\ln\frac{m^2}{m_2^2}-m_1^2\ln\frac{m^2}
{m_1^2}\right)\right],\end{equation}
\begin{equation}\label{66}a_n'=-2\delta_{l0}\left[\ln\frac{2\alpha}{n}+
\sum_{i=1}^n\frac{1}{i}+\frac{1}{2}-\frac{1}{2n}\right]+\frac{1-\delta_{l0}}
{l(l+1)(2l+1)},\end{equation} with $a_1'=-2\ln(2\alpha)-2$. The denominator 
$m_+m_-$ in (\ref{65}) will be discussed in the next section. The factor 
$1/m$ in front disappears in the expression for $E^2$, as $E^2\approx m^2
+2mE_b $ according to (\ref{1}).
\end{section}

\begin{section}{Bethe-Salpeter equation, vacuum polarization and form factor}
\label{ch7}
Continuing with the $T$-matrix of section 3, the two-photon exchange part 
$T^{(2)}$ may be used to derive corrections to the main interaction, 
including the Salpeter shift (\ref{65}). However, here we consider instead 
the more commonly used Bethe-Salpeter equation. It is formulated in the cms, 
where (\ref{51}) reduces to \beg{67}p_1^\mu=p^\mu+E_1g_{\mu 0},
\quad p_2^\mu=-p^\mu+E_2g_{\mu 0},
\end{equation} 
and $p^\mu$ is an integration variable. Suppressing the integration, the 
BS-equation applies to the two-fermion Greens function $G=G^{(16)}$
\beg{68}G^{(16)}=S^{(16)}+S^{(16)}K^{(16)}G^{(16)},\end{equation}
where $S^{(16)}$ is the product of two free fermion propagators in momentum 
space
\beg{69}S^{(16)}=(/\hspace{-0.23cm}p_1-m_1)^{-1}\otimes(/\hspace{-0.23cm}
p_2-m_2)^{-1}=(/\hspace{-0.23cm}p_1+m_1)\otimes(/\hspace{-0.23cm}p_2+m_2)/
D_1D_2,
\end{equation}
\beg{70} D_1=p_1^2-m_1^2+i\varepsilon=p^2+2E_1p^0+k^2+i\varepsilon,\quad D_2=
p_2^2-m_2^2+i\varepsilon=p^2-2E_2p^0+k^2+i\varepsilon,\end{equation}
and the kernel $K^{(16)}$ is the sum of all irreducible Feynman diagrams.
Our proposed eight-component formalism simplifies $S$ at the expense of $K$.
Taking $K$ from the matrix elements $\chi^\dagger K\psi$ analogous to the 
$8\times 8$ Matrix $M$ in (\ref{20}), one has 
\beg{71}S^{(16)}K^{(16)}=S^{(8)}K^{(8)},\quad S^{(8)}=\sum v_{LP}w^\dagger_{LP}
/D_1D_2.\end{equation}
In a matrix notation where $\gamma^0=\beta$ and $\gamma_5$ are identical for 
particles 1 and 2, the spin summation is now
\beg{72}\sum v_{LP}w^\dagger_{LP}=m_2E_1+m_1\beta E_2-m_-p^0-\gamma_5 \bld{p}
(m_2\bld{\sigma}_1-m_1\beta\bld{\sigma}_2).\end{equation}
It is only a first-order polynomial in $p_1,\;p_2$.
Consequently, the numerator of $S^{(8)}$ is linear in the integration
variables, whereas $S^{(16)}$ is quadratic.
One may in fact go one step further and use irreducible representations of the
Lorentz group, in which case one arrives at $S^{(4)}= 1/D_1 D_2$. This amounts
to the elimination of the dotted spinor components from $\psi$ \cite{BT}; it is
widely used in QCD calculations \cite{CS}. Four bound states, however, it has
the disadvantage of suppressing explicit parity invariance.

The last bracket in  (\ref{72}) is $(m_-\bld{\sigma}+m_+\Delta\bld{\sigma})/2$ 
in the notation (\ref{16}), and $m_-\bld{\sigma}$ may be converted into $m_+
\bld{\sigma}$ by the transformation (\ref{18})
\beg{73}v_{LP}=cv,\quad w_{LP}=c^{-1}w,\quad \sum vw^\dagger =(\epsilon-
\gamma_5\bld{\sigma}_1\bld{p}+\mu\beta+p^0m_-/m_+)m_+.\end{equation}
To handle the $p^0$-integration, one may use the formula 
$1/(x-i\varepsilon)=P/x
+i\pi\delta(x)$ ($P$=principal value):
\beg{74}2E/D_1D_2=2i\pi\delta(p^0)/(\bld{p}^2-k^2)-1/D_1(p^0+i\varepsilon)+
1/D_2(p^0-i\varepsilon),\end{equation}
and treat all $p^0$-integrals except the first one as perturbations of the 
Greens function. In these integrals, the spin summation (\ref{73}) reduces 
to $\sum
vw^\dagger / m_+=\epsilon-\bld{\alpha}\bld{p}+\mu\beta$, which is simply the 
expression for a single-particle spin summation. In the remaining integrals, 
the complete two-fermion propagator has the more explicit numerator
\beg{75}2E\sum vw^\dagger / m_+=E^2-(m_1^2+m_2^2-2m_1m_2\beta)(1-E p^0/m_+m_-)
-2\gamma_5\bld{p}_\rho\bld{\sigma}_1.\end{equation}
After integration, this $p^0$-dependence produces factors $(m_1^2+m_2^2)/m_+
m_-$ and $m_1m_2\beta/m_+m_-$. Such factors may also  arise from the 
denominators $1/D_1D_2^X$ of the crossed graph, which contain the 
combination $ p^0( E_1- E_2)=E p^0\hat m_+\hat m_-$.
To check the mass dependence in the $\delta_{l0}$-piece of $\Delta E_{Sal}'$
(\ref{65}), notice $m_2^2\ln (m^2/m_2^2)-m_1^2\ln (m^2/m_1^2)=m_+m_-\ln( 
m_1m_2/m^2)+(m_1^2+m_2^2)\ln(m^2/m_1m_2)$.

We conclude with a discussion of the potential $EV(r)\to V(\rho)$ in the 
presence of vacuum polarization and nuclear charge distribution. For a heavy 
particle 1 ($\mu^-$ or $\bar {p}$), electronic vacuum polarization is so 
large in low-$l$-states that it must be added to the Coulomb potential $V_C$, 
in the form of the Uehling potential:
\beg{84}V=V_C+V_U,\quad V_U=-\frac{Z\alpha^2}{3\pi r}
\int_0^\infty d\lambda^2e^{-\lambda r}S(\lambda^2),
\end{equation}
with $S_e=(1-4m_e^2/\lambda^2)^{1/2}(1+2m_e^2/\lambda^2)/\lambda^2
\Theta ( 4 m_e^2)$ in the 
electronic part of $V_U$. At fixed $\lambda$, the $r$-dependence is easily 
converted to a $\rho$-dependence, $(E/r)e^{-\lambda r}=\rho^{-1}e^{-\lambda 
E\rho}$:
\beg{85}V_U(\rho)=-\frac{Z\alpha^2}{3\pi\rho}\int_0^\infty 
d\lambda^2{\rm
e}^{-\lambda
E\rho}S(\lambda^2).\end{equation}
The substitution $\lambda E = \lambda '$ shows that the integral depends only
on $E^2$.
There is also a second-order recoil correction $V_{Ur}^{(2)}$ which is 
spin-independent \cite{FP}.
For antiprotonic atoms, the radius of the vacuum polarization cloud is 
much larger than the Bohr radius $(Z\alpha\mu)^{-1}$:  below a critical 
value $l_c$ of the orbital angular momentum $l$, vacuum polarization exceeds 
all relativistic effects (protonium $\bar{p}p$ and $\bar{p}\,^3$He 
have $l_c=3$ and $7$, respectively), for all values of $n$ \cite{BPS}. 
For these orbitals, $V$ is best constructed numerically from (\ref{85}), 
particularly for the calculation of annihilation which is localized at very 
small $\rho$. However, for the calculation of the fine
and hyperfine structure of these inner orbitals, the form (\ref{85}) 
suggests the introduction of a running electric coupling constant $\alpha_e$:
\beg{86}V=V_e+\delta V_U,\quad V_e=-Z\alpha_e/\rho,\quad\delta V_U
=V_U+Z(\alpha_e
-\alpha)/\rho,\quad \langle\delta V_U\rangle=0.\end{equation}In this manner, 
the nonperturbative result (3) remains applicable, with $\alpha_e>\alpha$.

Recoil corrections to an extended nuclear charge distribution $\rho_c(r)$ 
are particularly large for muonic atoms.
\beg{87}
V(\rho)=-Z\alpha\int d^3r|\bld{\rho}-\bld{r}/E|^{-1}\rho_c(r)
,\quad 
\rho_c(r)=\int d^3q {\rm e}^{-i\bld{qr}}F(q^2)
.
\end{equation}
Again, $V(\rho)$ is $E^2$-dependent. 
\end{section}

\begin{section}{angular momentum defects and Barker-Glover term}\label{ch8}
To order $\alpha_Z^2=Z^2\alpha^2$, $\delta l$ is easily found for particles 
of arbitrary masses $m_i$ and  $g$-factors $g_i = 2 ( 1 +\kappa_i ).\;
\delta l$ is conveniently expressed as follows:
\beg{88}\delta l/\alpha_Z^2 = -(1 + a) / (2 l +1 ).\end{equation}
With the abbreviations $\hat m_i=m_i/E,\;\hat\epsilon=\epsilon/E$ and the 
combinations \begin{eqnarray}c_1^2&=&1-\hat m_1^2-\hat\epsilon+2\kappa_1\hat 
m_2,\quad
c_2^2=1-\hat m_2^2-\hat\epsilon+2\kappa_2\hat m_1,\label{89}\\
&c^2&=1 + 2\kappa_1 {\hat m}_2 + 2 \kappa_2 {\hat m}_1 + 2\hat\epsilon(1-
g_1 g_2 /4),\label{90}\end{eqnarray}
one finds
\beg{91}a(l=f\pm 1)=\pm c^2/(2j+1)\pm g_1 g_2 \hat\epsilon/(4f+2)\end{equation}
(where a sign error of \cite{P95} for $l=f+1$ is corrected), and
\beg{92}a(l=f)=\pm\frac{c^2}{4F^2}\left[\sqrt{(2f+1)^2-16F^2\frac{c_1^2c_2^2}
{c^4}}\mp 1\right]
\approx\pm\frac{c^2}{2j+1}\mp\frac{2c_1^2c_2^2}{c^2(2f+1)}\left[1+
\frac{4F^2c_1^2c_2^2}{c^4(2f+1)^2}\right]
.\end{equation}
In (\ref{91}), $l=f\pm1$ implies $j=j_1=f\pm\kalb=l\mp\kalb$. For $g_1=g_2=2$, 
$c=1$ leads exactly to the Dirac fine structure component (\ref{42b}) of 
$\delta l$, for all values of $m_1$ and $m_2$, including $m_1=m_2$ as in 
positronium.
The case $l=f$ is more complicated, because the two values $j=l+\kalb$ and 
$j=l-\kalb$ are mixed by the hyperfine interaction. We therefore define $j$ 
as that value in the integer $2j+1$ which appears in the main term 
$c^2/(2j+1)$ of
$a(l=f)$ after expansion of the square root. 
It is frequently said that the fine structure contains the ``Barker-Glover'' 
term (for $l>0$) \cite{SY,Petal},
\beg{93}E_{BG}=\alpha_Z^4\mu^3(l-j)/[(2j+1)(2l+1)m^2n^3].
\end{equation}
This is true for a spinless nucleus, except that the electron's anomalous 
magnetic moment frequently reverses the effect \cite{P84}.
On the other hand, the weighted average over the hyperfine structure produces 
a more complicated expression,
\beg{94}\bar a=(l-j)[2c^2+g_1g_2\hat\epsilon/2-2(c_1^2c_2^2/c^2)
(1+4L^2c_1^2c_2^2/c^4(2l+1)^2)]/(2j+1).\end{equation}
This is the final result, and as a rule it is much larger than the 
Barker-Glover term. For $g_1=g_2=2$, however, $c_1^2c_2^2\approx\mu/m
\approx\hat\epsilon$ to this order in $\alpha$, and the main terms cancel 
in (\ref{94}). There remains a small rest,
\beg{95}\Delta\bar E=\alpha_Z^2\mu_{nr}{\bar\delta l}/(2l+1)n^3=-\mu_{nr}
\alpha_Z^4\bar a/n^3=4\mu_{nr}^3\alpha_Z^4(l-j)L^2/(2l+1)^3(2j+1)m^2n^3,
\end{equation}
which replaces $E_{BG}$ in the case of muonium.
\\

{\sc Acknowledgments:} The authors would like to thank S. Karshenboim and A. 
Yelkhovsky for helpful comments. This work has been supported by the 
Deutsche Forschungsgemeinschaft.
\end{section}



\begin{references}

\bibitem{GY} H. Grotch and D.R. Yennie, Rev. Mod. Phys. {\bf 41}, 350 (1969)


\bibitem{SY}J.R. Sapirstein and D. Yennie, in: Quantum Electrodynamics
(World Scientific, Singapore, 1990)


\bibitem{Petal} K. Pachucki et al, J. Phys. B {\bf 29}, 177 (1996)


\bibitem{Y}   A. Yelkhovsky, JETP {\bf 83}, 230 (1996)


\bibitem{CL} W.E. Caswell and G.E. Lepage, Phys. Lett. {\bf B 167}, 437 (1986)

\bibitem{L} P. Labelle, S.M. Zebarjad and C.P. Burgess, Phys. Rev. D {\bf 56},
            8053 (1997),
           
            A.H. Hoang, P. Labelle and S.M. Zebarjad, Phys. Rev. Lett. {\bf
            79}, 3387 (1997)
             


\bibitem{P92} H. Pilkuhn, J. Phys. B {\bf 25}, 289 (1992),

 H. Pilkuhn  and F. St\"audner, Phys. Letters A {\bf 178}, 156 (1993)


\bibitem{MP} M. Malvetti and H. Pilkuhn, Phys. Rep. C {\bf 248}, 1 (1994)


\bibitem{P95} H. Pilkuhn, J. Phys. B {\bf 28}, 4421 (1995)


\bibitem{Bre} E. Brezin, C. Itzykson and J. Zinn-Justin, Phys. Rev. D 
 {\bf 1}, 2349 (1970)

\bibitem{P84} H. Pilkuhn, J. Phys. B {\bf 17}, 4061 (1984)


\bibitem{gupta} S.N. Gupta, Quantum Electrodynamics (Gordon and Breach, New
 York 1977),\\
 S.N. Gupta, W.W. Repko and  C.J. Suchyta III, Phys. Rev. D 
 {\bf 40}, 4100 (1989)

\bibitem{PG} K. Pachucki and H. Grotch, Phys. Rev. A {\bf 51}, 1854 (1995),

             E.A. Golosov et al., JETP {\bf 80}, 208 (1995)

                

\bibitem{E} G.W. Erickson, J. Phys. Chem. Ref. Date {\bf 6}, 833 (1977)


\bibitem{K} S.G. Karshenboim, JETP {\bf 80}, 593 (1995)

\bibitem{BT} L.M. Brown, Phys. Rev. {\bf 111}, 957 (1958),
     
             M. Tonin, Nuo. Cim. {\bf 14}, 1108 (1959)

\bibitem{CS} G. Chalmers and W. Siegel, hep-ph/9708251

\bibitem{FP} J. Fr\"ohlich and H. Pilkuhn, J. Phys. B {\bf 17}, 147 (1984)

\bibitem{BPS} S. Barmo, H. Pilkuhn and H.G. Schlaile, Z. Phys. A {\bf 301},
 283 (1981)




\end{references}
\end{document}